\documentclass[12pt]{iopart}

\usepackage{graphicx}

\newcommand{\beq}{\begin{equation}}
\newcommand{\eeq}{\end{equation}}
\newcommand{\bk}{{{\bf{k}}}}

\newcommand{\bp}{{\bf{p}}}

\newcommand{\beqa}{\begin{eqnarray}}
\newcommand{\eeqa}{\end{eqnarray}}

\begin{document}
\title[Electronic topology of CoSi]{Band structure and unconventional electronic topology of CoSi}

\author{D A Pshenay-Severin$^1$, Yu V Ivanov$^1$, A A Burkov$^2$, A T  Burkov$^1$}

\address{$^1$Ioffe Institute, Saint Petersburg 194021, Russia}

\address{$^2$Department of Physics and Astronomy, University of Waterloo, Waterloo, Ontario 
N2L 3G1, Canada} 

\date{\today}
\begin{abstract}
Crystalline semimetals with certain space group symmetries may possess unusual electronic structure topology, distinct 
from the conventional Weyl and Dirac semimetals. 
Characteristic property of these materials is the existence of band-touching points with multiple (higher than two-fold) 
degeneracy and nonzero topological charge. 
CoSi is a representative of this group of materials exhibiting the so-called ``new fermions". 
We report on an ab initio calculation of the electronic structure of CoSi using density functional methods, taking into account the spin-orbit interactions.
We demonstrate the existence of band-touching nodes with four- and six-fold degeneracy, located at the $\Gamma$ and $R$ points in the first Brillouin zone and near the Fermi energy. 
We show that these band-touching points carry topological charges of $\pm 4$ and describe the resulting Fermi arc surface states, connecting the projections of these nodes onto the surface Brillouin zone. We also discuss the influence of many body $G_0W_0$ corrections on the electronic band structure and the topological properties of CoSi. 
\end{abstract}

\maketitle


\section{Introduction}
\label{sec:1}
Recent years have witnessed an explosion of interest in semimetals and metals with nontrivial electronic structure topology. 
The most well-known examples of these are Weyl and Dirac semimetals. 
The low-energy electronic structure of Weyl semimetals contains an even number of doubly-degenerate band-touching points, carrying topological charges $\pm 1$. 
In certain cases, the Weyl node pairs may exist at  the same position in the first Brillouin zone (BZ), producing Dirac points with four-fold degeneracy and zero topological charge. 
Weyl points with the same topological charge may also merge in crystals with certain symmetries, producing multi-Weyl 
semimetals~\cite{Fang2012}.

It has been demonstrated in Ref.~\cite{Bradlyn2016} that certain space groups allow the existence of nodes with three-, four- (but distinct from Dirac), six-, and eight-fold degeneracy and nonzero topological charge. 
These band-touching points were called ``new fermions" in Ref.~\cite{Bradlyn2016} in reference to the fact that such structures, unlike Weyl and Dirac nodes, do not exist in the relativistic quantum field theory context, being prohibited by Lorentz invariance. 

In this paper we show that CoSi is an example of such a ``new fermion" material. 
CoSi has been studied since the 1960-ies as a thermoelectric material (see review~\cite{Fedorov1995}). 
Early studies demonstrated that its transport properties could be explained if one assumed that the valence and conduction bands overlapped by about 20-40 meV~\cite{Fedorov1995,Asanabe1964}.
The band dispersions were assumed to be parabolic with the effective masses of electrons and holes of $2 m$ and $4-6 m$, where $m$ is the free electron mass~\cite{Fedorov1995,Asanabe1964}. 
However, as will be demonstrated in this work, the real electronic structure of CoSi is significantly more complex. 

The crystal structure of CoSi does not contain an inversion center and belongs to the space group 198 ($P2_13$). 
A simple cubic unit cell contains four formula units and is characterized by the parameters $a_0 = 4.4445$ \AA, 
$x_{\mathrm{Co}} = 0.144$, and $x_{\mathrm{Si}} = 0.846$~\cite{Fedorov1995,Zelenin1964}.
The BZ is a cube with the $\Gamma$-point at the center of the cube, $R$-points at the vertices, $X$ at the centers of the faces and $M$ at the centers of the edges. 
Electronic structure calculations using density functional methods were done in Ref.~\cite{Pan2007,Sakai2007} without taking into account the spin-orbit (SO)
coupling and in Ref.~\cite{Ishii2014} with the SO coupling included. 
Symmetry analysis showed the existence of six-fold degenerate band-touching nodes at the $R$ point in the BZ and a ${\bk \cdot \bp}$ 
Hamiltonian, describing the electronic structure in the vicinity of the $R$ point, was derived~\cite{Bradlyn2016,Manes2012}.
The influence of the SO interactions on the electronic structure was discussed in Ref.~\cite{Ishii2014}.

The compounds, isostructural to CoSi, are also known. These include CrSi, MnSi and RhSi with metallic-type conductivity, while FeSi, RuSi and OsSi were classified as narrow-gap semiconductors. 
Recent calculations of the electronic spectrum of RhSi and CoSi \cite{Chang2017,tang2017} revealed 4 - and 6 - fold band degeneracies at the $\Gamma$ and $R$ points of the Brillouin zone, respectively. 
It was also shown that Chern numbers of these nodes are equal to $\pm 4$, and the length of each Fermi arc exceeds dimensions of the surface Brillouin zone.

This work was initiated by our experimental results on low-temperature resistivity and thermopower of CoSi and Co$_{1-x}$Fe$_x$Si ($x$=0.04)~\cite{burkov2017a_tr}. 
The properties show unusual for metal temperature variation which probably can be linked to topological features of CoSi electronic structure.
Here we present an ab initio calculation of the electronic structure of CoSi, 
focusing on the vicinity of the $\Gamma$ and $R$ points in the BZ. 
We show that the band degeneracy nodes at these points carry nontrivial 
topological charges $\pm4$. A special attention is payed to origin  
of this large topological charge. Since it is known that nodes with non-linear dispersion can have topological charge larger than unity~\cite{Huang2016}, we used effective linearized $\bk\cdot\bp$ Hamiltonian to show that large topological charges of these nodes at $\Gamma$ and $R$ points are not connected to a non-linear dispersion. 
We also confirm that these non-trivial topological charges lead to the formation of Fermi  
arc surface states in the two-dimensional surface BZ.

It is well known that electron-electron correlations in transition metal compounds can play a very important role. 
Recently we have demonstrated that the inclusion of many body corrections significantly improves the theoretical description of the Seebeck coefficient of CoSi in the constant relaxation time approximation~\cite{PshenayICT2017}.
This suggests that these corrections may have a significant impact on topological properties and the shape of the Fermi arcs.
Therefore here we also examine the effect of many body $G_0W_0$ corrections \cite{Shishkin2006,Shishkin2007} on the quasiparticle band structure, topological charges and Fermi arcs.

\section{Electronic structure of cobalt monosilicide without $G_0W_0$ corrections}
\label{sec:2}
\begin{figure*}
\centering
\includegraphics[width=0.9\textwidth]{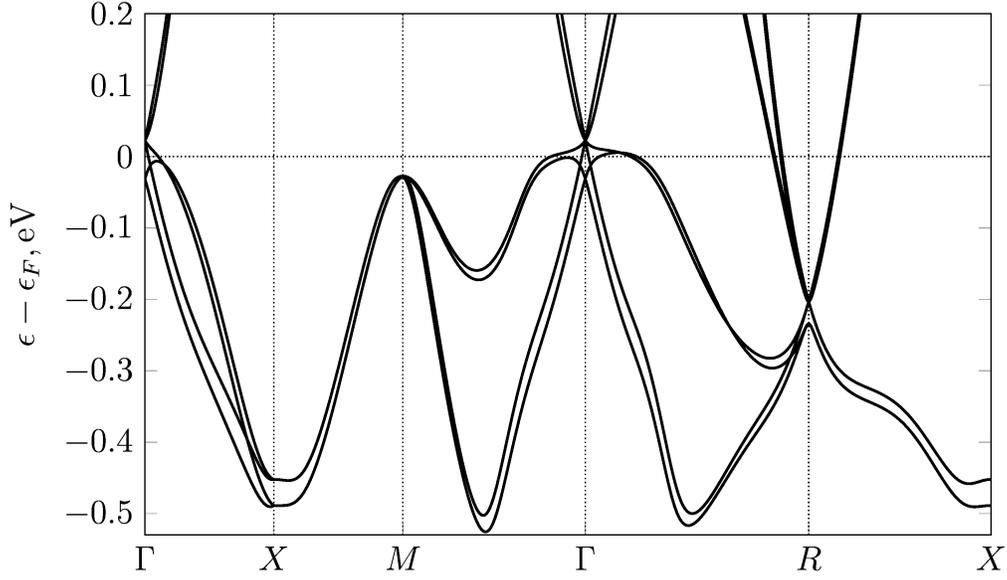}
\caption{\label{fig:bs1}The band structure of CoSi near the Fermi level calculated with the inclusion of spin-orbit coupling.}
\end{figure*}

We calculate the electronic structure of CoSi within generalized gradient approximation (GGA-PBE), taking into account the SO interactions, with the help of the VASP density functional package~\cite{Kresse1996,Kresse1999}.
We used optimized lattice parameters $a_0 = 4.430$ \AA, $x_{\mathrm{Co}} = 0.145$, and $x_{\mathrm{Si}} = 0.843$ (lattice relaxation parameter was 1~meV/\AA), which agree well with the experimental values~\cite{Fedorov1995,Zelenin1964}. 
The calculations were performed on $8 \times 8 \times 8$  Monhorst-Pack mesh with 350~eV cutoff energy. 
In agreement with~\cite{Ishii2014,Pan2007,Sakai2007,tang2017,PshenayICT2017} we find that the states near the Fermi energy are located near the $\Gamma$, $M$ and $R$ points in the BZ (see Fig.~\ref{fig:bs1}). 
While the band dispersion at the $M$ point is parabolic, there are linear band crossings at the $R$ and $\Gamma$ points.

\begin{figure*}
\centering
\includegraphics[width=0.94\textwidth]{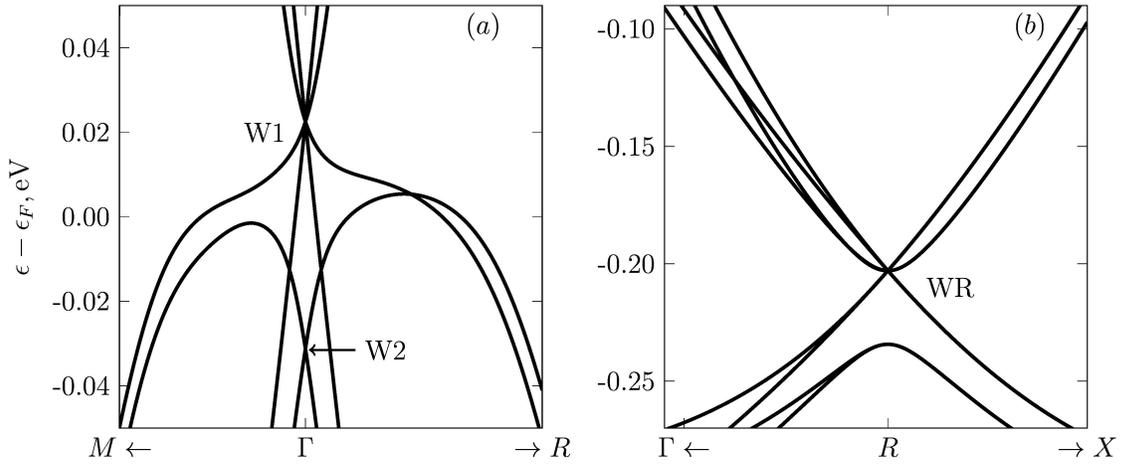}
\caption{\label{fig:bs_zoom}Band structure of CoSi near $\Gamma$ (\textit{a}) and $R$ (\textit{b}) points.}
\end{figure*}

Fig.~\ref{fig:bs_zoom}  shows the part of the electronic structure of CoSi near the $\Gamma$ and $R$ points in the BZ. 
It can be shown that the degenerate multiplets at these points arise due to the crystallographic symmetry as well as 
the time reversal symmetry. 
Indeed, if SO interactions are neglected, the bands shown in Fig.~\ref{fig:bs_zoom}$a$ form a spin-degenerate orbital triplet, transforming under the irreducible representation $\Gamma_4$ of the little group of the $\Gamma$ point. 
SO interaction splits the orbital triplet~\cite{Sakai2007} into a doublet $\bar \Gamma_5$ (Weyl node W2 in Fig.~\ref{fig:bs_zoom}$a$) and two degenerate doublets $\bar \Gamma_6$ and $\bar \Gamma_7$ (node W1). 
Irreducible representations $\bar \Gamma_6$ and $\bar \Gamma_7$ are conjugated. 
The bases of these representations are transformed into each other under time reversal, guaranteeing their degeneracy.
This leads to the appearance of the node W1 at the $\Gamma$ point, which looks like a pair of Weyl fermions with different 
Fermi velocities, sharing the same origin in momentum space (this is not an ordinary Dirac point). 

\begin{figure*}
\centering
\includegraphics[width=0.99\textwidth]{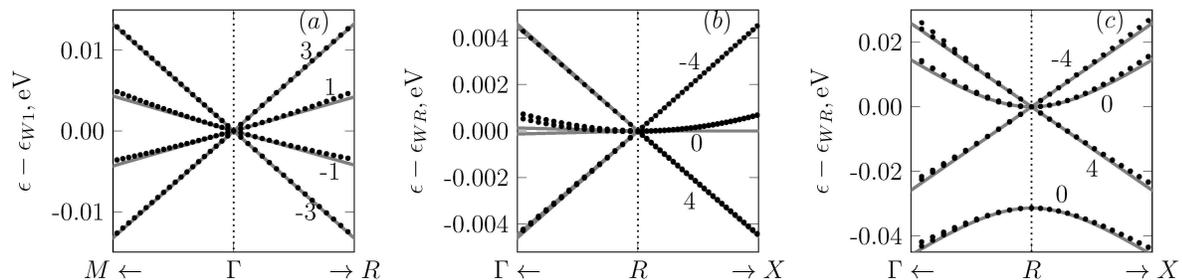}
\caption{\label{fig:bsGR} Energy dispersions of fermions near $\Gamma $($a$) and $R$ ($b$, $c$) points. The solid lines represent results of calculations based on the linearized $\mathbf{k}\cdot \mathbf{p}$ Hamiltonians ($a$ - Eq.~(\ref{eq:2}); $b$ - Eq.~(\ref{eq:5}), $c$ - Eqs.~(\ref{eq:7})-(\ref{eq:8})), the points represent results of \textit{ab initio} calculations of the band structure. Chern numbers are shown close to corresponding bands. Energies are measured from the band crossings.}
\end{figure*}

At linear order in momentum, the $\bk\cdot\bp$ Hamiltonian, which is invariant under transformations of the little group of the $\Gamma$ point and time reversal takes the form:
\begin{equation}
\label{eq:1}
\fl
H_{\Gamma} = \left(
\begin{array}{cccc}
a k_z & a (k_x - i k_y) & b (\nu_3 k_x - \nu_6 k_y) & b k_z \\
a (k_x +  i k_y) & - a k_z & b k_z & - b (\nu_3 k_x + \nu_6 k_y) \\
b^* (\nu_3^* k_x - \nu_6^* k_y) & b^* k_z & -a k_z & -a (k_x + i k_y) \\
b^* k_z & -b^* (\nu_3^* k_x + \nu_6^* k_y) & -a (k_x - i k_y) & a k_z
\end{array}
\right).
\end{equation}
Here $\nu_3=e^{i \pi /3}$,  $\nu_6=e^{i\pi /6}$, $\bk$ is the (dimensionless, in units of the primitive translations of the reciprocal lattice) wavevector, measured from the $\Gamma$ point, $a$ is a real and $b$ is a complex parameter. 
When deriving this Hamiltonian, the matrices of the irreducible representations  $\bar \Gamma_6$ and $\bar \Gamma_7$  presented in the Bilbao Crystallographic Server \cite{Aroyo2011,Elcoro2017} were used.
The energy eigenvalues of this Hamiltonian are given by
\beq
\label{eq:2}
\epsilon_{\pm}^2(\bk) = \bk^2 [a^2 + |b|^2 \pm \sqrt{(a^2 + |b|^2)^2 - \textrm{det} H_{\Gamma}/\bk^4}]. 
\eeq
The parameters $a$ and $b$ are found by fitting Eq.~\eref{eq:2} to the ab initio spectrum in Fig.~\ref{fig:bs_zoom}$a$. 
We obtain $a = \pm0.56\mathrm{eV}$ and $|b| = 1.19\mathrm{eV}$.
The eigenvalues do not depend on $\mathrm{arg}(b)$. 
The fitting near the $\Gamma$ point is shown in Fig.~\ref{fig:bsGR}$a$. 

At the $R$ point in the BZ we find a novel six-fold degenerate node, predicted in Ref.~\cite{Bradlyn2016}. 
This node is located below the Fermi energy. A little lower in energy we also find a doublet, split from the 
six-fold degenerate multiplet by the SO interaction. 
Bands, which form the multiplet, are doubly-degenerate at boundary of the BZ due to the presence of two-fold screw 
rotations in the crystal symmetry group and time reversal symmetry. 
In other words, the faces of the cubic unit cell in momentum space form a Weyl nodal surface~\cite{Bradlyn2016}.
Among the irreducible double-valued representations of the little group of $R$, the highest-dimensional representation is the real three-dimensional representation $\bar R_7$. 
Time reversal symmetry doubles this dimensionality to six~\cite{Elcoro2017}.
The corresponding band Hamiltonian near the $R$ point to linear order in momentum is given by~\cite{Bradlyn2016}
\beqa
\label{eq:3}
H_R = \left(
\begin{array}{cc}
H_{R_7}(\tilde{a}) & \tilde{b} H_{R_7}(1) \\
\tilde{b}^* H_{R_7}(1) & -H_{R_7}^{*}(\tilde{a})
\end{array}
\right),
\eeqa
where
\beqa
\label{eq:4}
 H_{R_7}(\tilde{a}) = \left(
 \begin{array}{ccc}
 0 & \tilde{a} k_x & \tilde{a}^* k_y \\
 \tilde{a}^* k_x & 0 & \tilde{a} k_z \\
 \tilde{a} k_y & \tilde{a}^* k_z & 0
 \end{array}
 \right), 
 \eeqa
 and $\tilde{a}$ and $\tilde{b}$ are complex parameters. 
 Wave vector is measured from the $R$ point.
 Band dispersions may be found from the secular equation
 
 \beq
 \label{eq:5}
 {{\varepsilon }^{2}}{{\left[ {{\varepsilon }^{2}}-v^2\,{{k}^{2}} \right]}^{\,2}}+\det {{H}_{R}}=0,                                            
 \eeq
where $\textrm{det} H_R = - 4 (k_x k_y k_z)^2 (v^2 - u^2) (v^2-4 u^2)^2$, $v = (|\tilde{a}|^2 + |\tilde{b}|^2)^{1/2}$ and $u = |\tilde{a} \sin(\arg(\tilde{a}))|$.

If one of the components of $\bk$ is equal to zero, i.e. $\bk$ is on the cubic BZ boundary, $\textrm{det} H_R$ 
vanishes. 
In this case, the solutions of Eq.~\eref{eq:5} take a simple form 
\beq
\label{eq:6}
\epsilon_0 = 0, \,\, \epsilon_{\pm} = \pm v k, 
\eeq
where all the branches are two-fold degenerate. 
From the slope of the linearly-dispersing bands in these planes one can determine $v = 1.3 \mathrm{eV}$. Along the $R$ -- $\Gamma$ direction the linearly-dispersing bands split into two branches with the slopes $1.32 \mathrm{eV}$ and $1.28 \mathrm{eV}$ that allowed to obtain $\textrm{det} H_R=(k_x k_y k_z)^2 \, 0.124\,\mathrm{eV}^6$. The solutions of equation (\ref{eq:5}) with these parameters are shown in Fig.~\ref{fig:bsGR}$b$.

The determinant $\textrm{det} H_R$, and thus the roots of the secular equation, depend only on $|\tilde{a}|$, $|\tilde{b}|$ and $\textrm{arg}(\tilde{a})$, but not on $\textrm{arg}(\tilde{b})$. The parameters of Hamiltonian (\ref{eq:3}) cannot be determined uniquely. Using fitted values of $v$ and $\textrm{det} H_R$, one obtaines $u = $ 0.62~eV, 0.68~eV or 1.29~eV. For each of these three values the phase of $\tilde{a}$ can be taken from the range $\arg(\tilde{a}) \in [\alpha_0 + \pi m, \pi-\alpha_0 + \pi m]$, where $\alpha_0 = \arcsin(u/v)$ and $m$ is an integer (because the band dispersions are periodic functions of $\textrm{arg}(\tilde{a})$ with period $\pi$). Fixing $\arg(\tilde{a})$, then allows us to determine $|\tilde{a}|$ and $|\tilde{b}|$. The calculation of topological charges, discussed below, shows that this ambiguity does not affect their absolute values at fixed $m$, but their signs are opposite for odd and even $m$.

The appearance of chiral fermions at high-symmetry points in the BZ, in particular in compounds with space group 198, 
has also been discussed in Ref.~\cite{Manes2012}. 
It was demonstrated that in the absence of the SO interactions, ignoring the spin and time reversal symmetry, 
the $R$ point must host degenerate doublets (``orbital Weyl nodes"), which transform under irreducible representations 
$R_{1}$, $R_{2}$ or $R_{3}$ of the little group at $R$. 
Time reversal symmetry doubles the degeneracy and the number of ``orbital Weyl cones". 
In other words, the band dispersion in the vicinity of the $R$ point consists of pairs of ``Weyl" cones, distinguished by 
a ``valley" index. 
In addition, each cone has double spin degeneracy. 
Once SO interaction is included, it partially lifts the degeneracy, which results in the following band dispersions near the $R$ point~\cite{Manes2012}
\beq
\label{eq:7}
\epsilon_{1 \pm} = \frac{\Delta}{4} \pm v k, 
\eeq
and 
\beq
\label{eq:8}
\epsilon_{2 \pm} = - \frac{\Delta}{4} \pm \sqrt{\left(\frac{\Delta}{2}\right)^2 + v^2 \bk^2}, 
\eeq
where $v$ is the same velocity as in (\ref{eq:6}) and $\Delta$ is the strength of the SO interactions. 
We note that the ``valley" degeneracy still remains. 
Eq.~\eref{eq:7} describes massless fermions with linear dispersion, while Eq.~\eref{eq:8} --- massive Dirac particles. 
The spectrum branches $\epsilon_{1\pm}$ and $\epsilon_{2+}$ are degenerate at the $R$ point. 
In a neighborhood of the $R$ point, Eqs.~\eref{eq:7} and \eref{eq:8}, can be fitted to ab initio spectrum, giving 
$\Delta=$0.031~eV with the same $v$ as in (\ref{eq:6}) (see Fig.~\ref{fig:bsGR}$c$).
Eqs.~\eref{eq:5} and \eref{eq:7}, \eref{eq:8} complement each other. 
Dispersion relations, obtained by solving Eq.~\eref{eq:5}, shown in Fig.~\ref{fig:bsGR}$b$, take into account the lifting of band degeneracy 
away from the $R$ point, while Eqs.~\eref{eq:7} and \eref{eq:8} describe the band splitting, induced by the SO interactions, and 
parabolicity of some of the branches in the vicinity of the $R$ point (Fig.~\ref{fig:bsGR}$c$). 

The band structure shown in Figs.~\ref{fig:bs1} to \ref{fig:bsGR} hosts also numerous band crossings along the high-symmetry $\Gamma-X$ and $\Gamma-M$ lines, which exist due to the lack of inversion symmetry in the crystal structure of CoSi. As shown in Ref.~\cite{tang2017}, these crossings are ordinary type-I and type-II Weyl points. Here we focus only on non-Weyl ("new fermion") nodes at the time-reversal invariant points in the BZ.

\begin{figure*}
\centering
\includegraphics[width=0.49\textwidth]{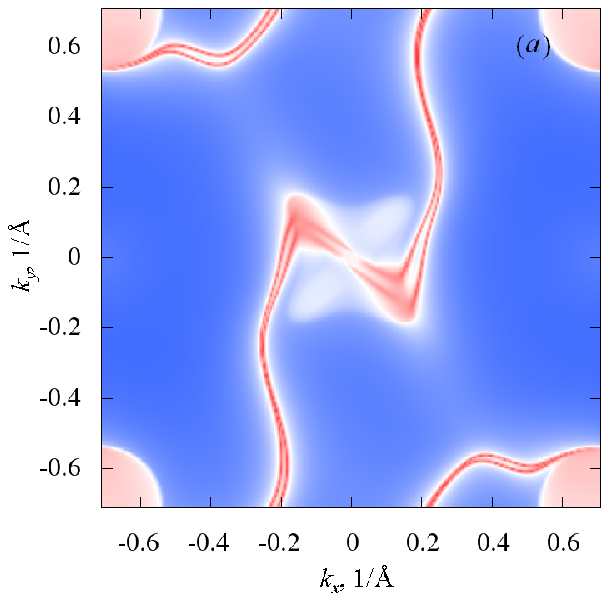}
\includegraphics[width=0.49\textwidth]{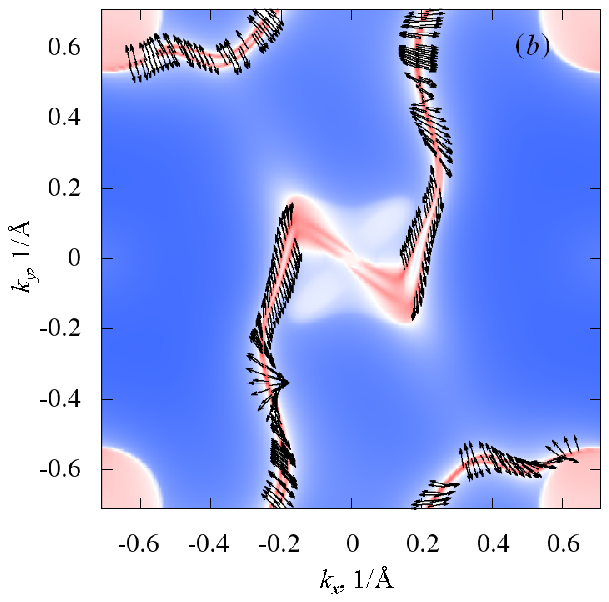}
\caption{\label{fig:arc} Distribution of the density of states in the (001) surface Brillouin zone of CoSi slab ($a$) and the spin texture of surface states ($b$). The $\Gamma$ point of the bulk Brillouin zone is projected to the square center $\overline{\Gamma }$, the points $R$ and $M$ are projected to the square vertices $\overline{M}$. }
\end{figure*} 

\section{Chern numbers and Fermi arcs}
\label{sec:3}
One may speak of nontrivial band topology of CoSi if the nodal points, described in the previous section, carry nontrivial 
topological charges, which in turn may lead to the appearance of Fermi arc surface states. 
Near the $\Gamma$ point, the band degeneracy is absent for small but finite wave vectors, and the topological charge is equal to the flux of the Berry curvature $\mathbf{\Omega}(\bk)$ through a closed surface 
in momentum space, enclosing a given degeneracy point. 
We evaluate the Berry curvature of the band $n$ as~\cite{Shen2012,Baggio2017,Resta}
\beqa
\label{eq:9}
\Omega^{(n)}_{\alpha}(\bk) = - \epsilon_{\alpha \beta \gamma} \textrm{Im}&\sum_{n' \neq n} \frac{\langle u_n(\bk)|H'_{\beta}(\bk)|u_{n'}(\bk) \rangle \langle u_{n'}(\bk)|H'_{\gamma}(\bk)|u_n(\bk) \rangle}{(\epsilon_n(\bk) - \epsilon_{n'}(\bk))^2},
\eeqa
where $|u_n(\bk) \rangle$ and $\epsilon_n(\bk)$ are the eigenstates and energy eigenvalues of the Hamiltonian $H(\bk)$, 
$H'_{\beta}(\bk) = \partial H(\bk)/ \partial k_{\beta}$, and $\epsilon_{\alpha \beta \gamma}$ is the fully antisymmetric tensor. 
One may calculate the Berry curvature using either the first-principles Hamiltonian, or the $\bk \cdot \bp$ Hamiltonian~\eref{eq:1}. 
In the former case, the Hamiltonian at an arbitrary point in the BZ is calculated as follows. 
Using the wavefunctions, obtained in VASP, we utilize the Wannier90 package~\cite{Mostofi2014} to construct a tight-binding Hamiltonian.
After that, the momentum-space Hamiltonian $H(\bk)$ was found by lattice Fourier transform~\cite{Wu2017,z2pack,z2pack2}, for which we used the TBModels package~\cite{z2pack}.
As a result of these calculations, we find that at the W1 node, the topological charges of the two pairs of touching bands are equal to $\pm3$ and $\pm1$, see Fig.~\ref{fig:bsGR}$a$. 
The W2 node is a standard Weyl node with the Chern number $1$. 
We have also checked our results for the W1 node using the linearized bands Hamiltonian Eq.~\eref{eq:1}. 
In this case, the signs of the topological charges depend on the sign of the parameter $a$ and do not depend on $\textrm{arg}(b)$. 
They agree with the charges, found from the tight-binding Hamiltonian, when $a > 0$ is taken.    

The above procedure for calculating the topological charges is inapplicable for states near the $R$ point. This $k$-point is located at the crossing of cube faces, where bands are pairwise degenerate. Therefore, the non-Abelian Berry curvature should be used~\cite{Baggio2017,Resta}, and one can calculate only the total topological charge for a set of degenerate bands, 
which is enough for our purposes. It is convenient to denote the indices
 of these bands as $\{n\}$. For the calculation of non-Abelian Berry curvature, equation (\ref{eq:9}) should be corrected by excluding from summation over ${n}'$ all bands, degenerate with $n$-th one (i.e. by summing over $n'\notin \{n\}$). After that, the non-Abelian Berry curvature for a group of degenerate bands can be obtained by summing up $\Omega _{\alpha }^{(n)}(\mathbf{k})$ over $n\in \{n\}$. As a result we find that, using the first-principles Hamiltonian, the total topological charge of the two linearly dispersing bands is equal to $-4$ and $+4$, respectively (see Fig.~\ref{fig:bsGR}$b$). The parabolic bands have zero Chern numbers. The same topological charges were obtained from the low-energy Hamiltonian Eq.~\eref{eq:3}. 
The charges have the same absolute values for all $\arg(\tilde{a}) \in [\alpha_0 + \pi m, \pi-\alpha_0 + \pi m]$ at fixed $m$. But, similarly to the case of the Chern numbers at the $\Gamma$ point, they change their signs when $m$ switches from odd to even.
 Topological charges are independent of the phase of the complex parameter $\tilde{b}$.
Interestingly, we obtain the same topological charges using the Hamiltonian from Ref.~\cite{Manes2012}, in spite of the fact that in the neighborhood of the $R$ point it describes pairwise degenerate bands in all direction of the BZ (see Fig.~\ref{fig:bsGR}$c$). But in this case, unlike in the case of the regular Dirac cones, the contributions of both degenerate bands from each pair to the topological charge are of the same sign. 

Therefore, the total topological charge of the two multiplets at the $\Gamma$ and $R$ points is equal to zero, in agreement with 
the Nielsen-Ninomiya theorem~\cite{Nielsen1981}.
We note that multiple degeneracy of the bands leads to higher topological charges of some of the bands, in spite of their 
linear dispersions.
Similar results for RhSi have been obtained in Ref.~\cite{Chang2017}.
This feature distinguishes CoSi from multi-Weyl semimetals, such as SrSi$_2$~\cite{Huang2016}, which possess higher Chern numbers due to band dispersion nonlinearity~\cite{Fang2012}.
\begin{figure*}
\centering
\includegraphics[width=0.8\textwidth]{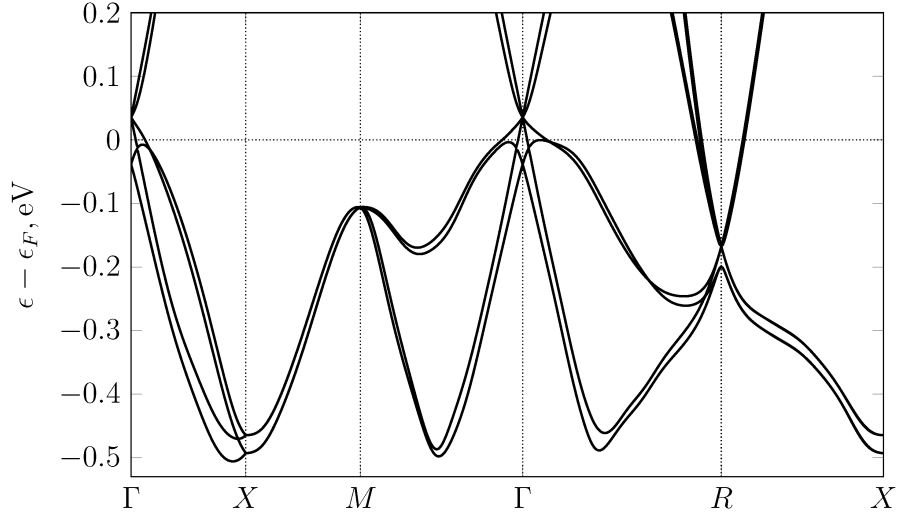}
\caption{\label{fig:bs1GW}The band structure of CoSi near the Fermi level calculated with the inclusion of spin-orbit coupling and $G_0W_0$ corrections.}
\end{figure*}

\begin{figure}
\centering
\includegraphics[width=0.49\textwidth]{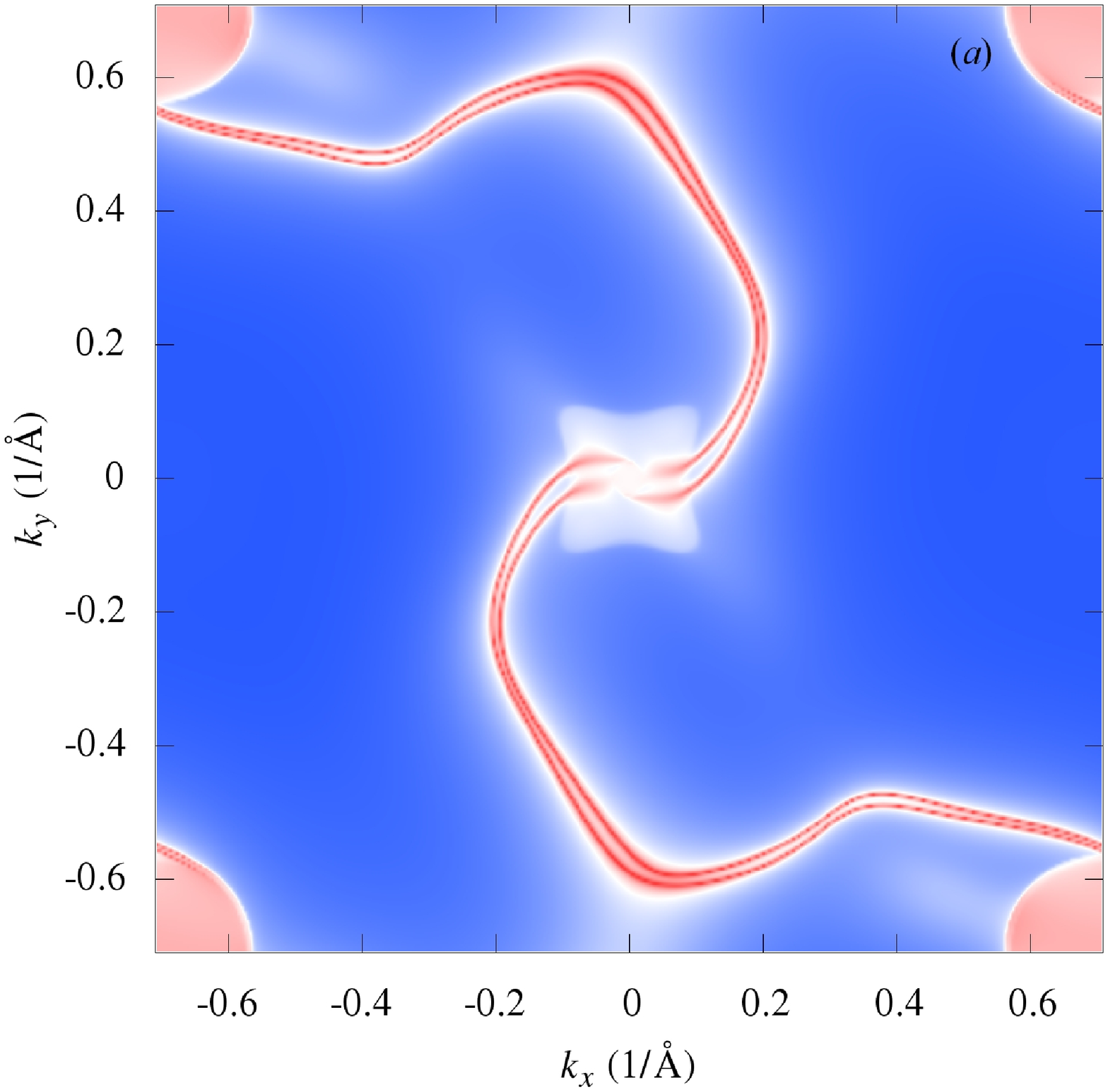}
\includegraphics[width=0.49\textwidth]{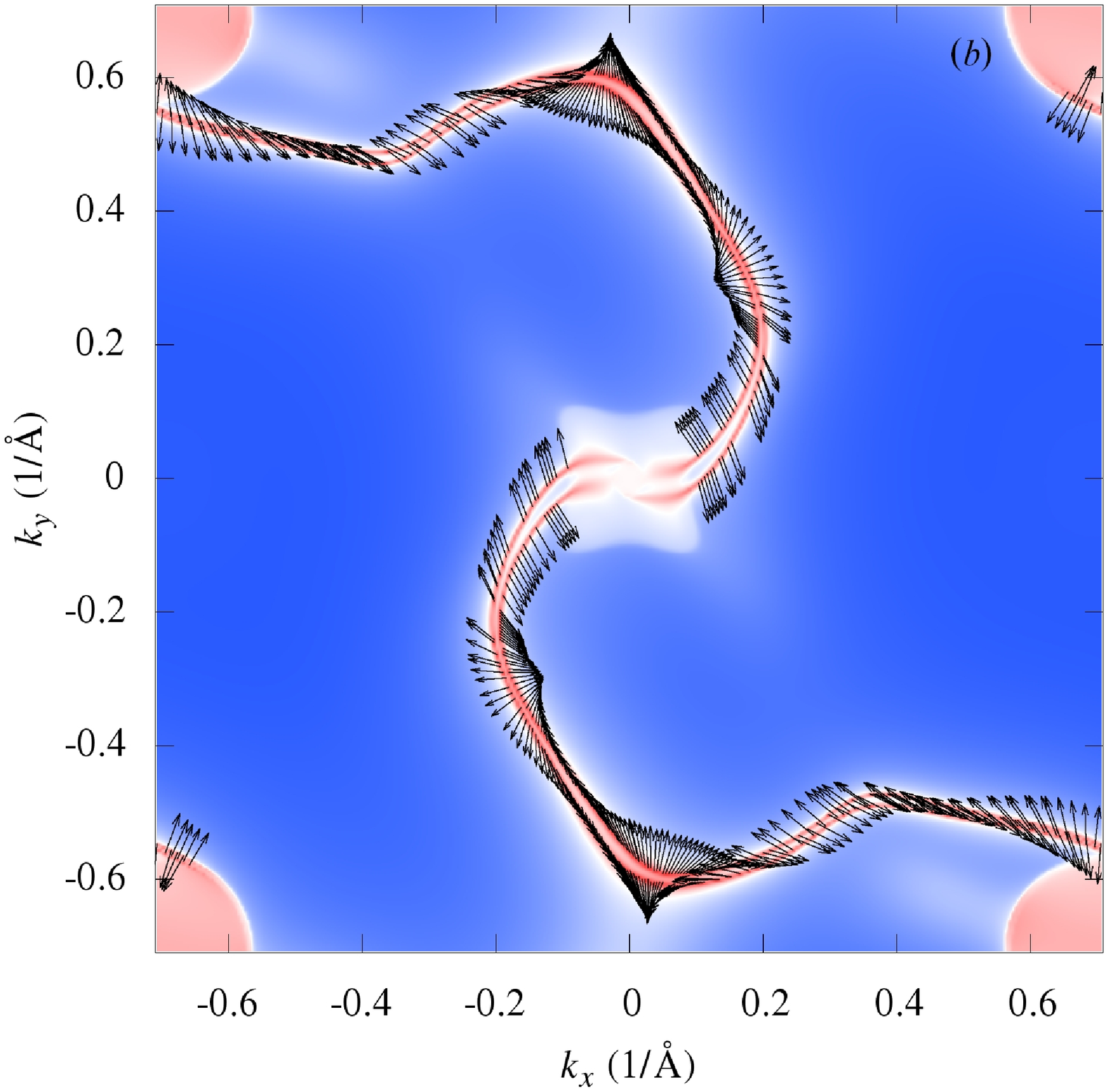}
\caption{\label{fig:arcGW} Distribution of the density of states in the (001) surface Brillouin zone of CoSi slab ($a$) and the spin texture of surface states ($b$) calculated in $G_0W_0$ approximation (see notation in Fig.~\ref{fig:arc}).
}
\end{figure} 
Due to the nonzero topological charges of the nodes at the $\Gamma$ and $R$ points, we expect Fermi arc states to exist 
on certain sample surfaces.
We calculated the surface density of states using the WannierTools package~\cite{Mostofi2014,Lopez1985}.
Fig.~\ref{fig:arc}$a$ shows the density of states at the Fermi energy in the two-dimensional BZ of the $(001)$ surface. 
The corresponding picture for the opposite surface of the slab is obtained by the rotation by $\pi $ around $k_x$ or $k_y$ axis. 
The spin texture of the surface states is shown in the Fig.~\ref{fig:arc}$b$. 
One can see that there are four Fermi arcs, which begin and end near the $\bar \Gamma$ and $\bar M$ points, 
which correspond to the center and the corners of the surface BZ. 
The bulk multiplets, described above, all have different energy and the Fermi level in undoped CoSi is situated in between. 
The deviation of the W1 point from the Fermi energy is small (0.023~eV), while that of the $R$-point multiplet is larger (-0.2~eV). 
As a result, the Fermi arcs in Fig.~\ref{fig:arc} emerge essentially from the center of the surface BZ, where the projection of the small hole pocket (with a linear dispersion) of the bulk Fermi surface with a large topological charge is located. The Fermi arcs end at the circles around the $\bar M$ points, which represent projections of the electronic pockets of the bulk Fermi surface onto the two-dimensional BZ.

\section{The influence of $G_0W_0$ corrections on topological properties} \label{sec:3_1}

Since calculations with the  $G_0 W_0$ corrections are computationally very demanding they were performed using the experimental lattice parameters~\cite{Fedorov1995, Zelenin1964}.
The calculations were made using the VASP software package~\cite{Shishkin2006,Shishkin2007} with the same parameters of the Monkhorst-Pack grid and the energy cutoff as quoted above. The interpolation of the band structure was made using the Wannier90 package~\cite{Mostofi2014}.

The band structure, calculated with the $G_0 W_0$ correction, is shown in the Fig.~\ref{fig:bs1GW}. One can see that compared to the Fig.~\ref{fig:bs1} the band maximum at the $M$-point is shifted downward in energy by about 0.1~eV. 
The dispersion of the heavy holes at the $\Gamma $ point becomes significantly stronger. 
At the same time, the spectrum near the $R$ point changes only slightly.
The difference in energy between the nodes at the $\Gamma$ and $R$ points does not change significantly. 
We evaluated the Berry curvature using the first principle Hamiltonian, following the same procedure as for the case without many body corrections and obtained the same result. 
As the appearance of linear band crossing is the result of the specific crystallographic symmetry of the 198 space group, it is not surprising that the Chern numbers, calculated for this case, are the same as in the standard PBE approximation. On the other hand, the shape of the Fermi arcs does change noticeably. The Fermi arcs and the spin texture, obtained with the $G_0W_0$ corrections using WannierTools~\cite{Mostofi2014,Lopez1985}, are shown in the Fig.~\ref{fig:arcGW}. It can be seen that the Fermi arcs become more compact compared with the Fig.~\ref{fig:arc} and completely reside inside the first Brillouin zone. This may be connected with the change of the band dispersion and also with the shift of Fermi level relative to node energies at the $\Gamma$ and $R$ points.
We note, that the shapes of the Fermi arcs obtained in this work (both, with and without the $G_0 W_0$ corrections) and in Refs.~\cite{tang2017,Chang2017} are all noticeably different from each other.
The reason for this discrepancy is not clear at the moment.

\section{Conclusions}\label{sec:4}
The  calculations, reported in this paper, demonstrate that CoSi is a novel type of topological metal, 
whose electronic structure is significantly different from ordinary Weyl, Dirac and even more exotic multi-Weyl and double-Dirac semimetals. 
The main new feature of the electronic structure of CoSi is the presence of topologically-nontrivial band-touching 
nodes with multiple band degeneracy (``new fermions").
The nodes are located at two time reversal invariant points in the BZ and carry nonzero topological charges. 
The multiplet, located at the $\Gamma$ point is four-fold degenerate, while the node at the $R$ point is six-fold 
degerate. The two multiplets carry total topological charges of magnitude 4 and opposite signs, in spite of band linearity. 
Both multiplets are located not far from the Fermi energy and are separated by about 220~meV in energy. 
The resulting surface states form four Fermi arcs, which begin and end near the projections of the bulk $\Gamma$ and $R$ points onto the surface BZ.
The calculation using the many body $G_0W_0$ corrections to electronic band structure shows that the topological charges at 
$\Gamma$ and $R$ points are insensitive to them, as expected, but the shape of the Fermi arcs does significantly depend on the approximations used due to the change in the band dispersions and the Fermi level position.

\section*{Acknowledgments}\label{sec:5}
We acknowledge financial support by the Russian Science Foundation, project no. 16-42-01067.

\section*{References} 
\bibliography{bibl}

\end{document}